\documentclass[12pt]{article}
\usepackage{amssymb,amsmath}
\usepackage{comment}
\usepackage{graphicx}
\usepackage{subfigure}
\usepackage{hyperref}

\begin{document}

\thispagestyle{empty}
\begin{flushright}
NCTS-TH/1504 
\end{flushright}
\vskip2cm
\begin{center}
{\Large AdS$_{2}$/CFT$_{1}$, Whittaker vector and Wheeler-DeWitt equation}

\vskip2cm
Tadashi Okazaki\footnote{tadashiokazaki@phys.ntu.edu.tw} 

\bigskip\bigskip
\vskip3mm
{\it 
Department of Physics and Center for Theoretical Sciences,\\ 
National Taiwan University, Taipei 10617, Taiwan
}
\end{center}

\vskip2cm
\begin{abstract}
We study the energy representation of 
conformal quantum mechanics as the Whittaker vector 
without specifying classical Lagrangian. 
We show that a generating function of 
expectation values among two excited states of 
the dilatation operator in conformal quantum mechanics 
is a solution to the Wheeler-DeWitt equation 
and it corresponds to the AdS$_{2}$ partition function  
evaluated as the minisuperspace wave function in Liouville field theory. 
We also show that the dilatation expectation values 
in conformal quantum mechanics 
lead to the asymptotic smoothed counting function of 
the Riemann zeros. 
\end{abstract}


\newpage
\setcounter{tocdepth}{2}

The holographic principle \cite{'tHooft:1993gx,Susskind:1994vu} 
states that quantum gravity on $(d+1)$-dimensional manifold 
can be described by a theory on its $d$-dimensional boundary. 
The AdS$_{d+1}$/CFT$_{d}$ correspondence 
\cite{Maldacena:1997re} which is one of the greatest productions of
string theory 
provides the most successful realization as the relationship 
between effective gauge theories of the brane dynamics 
and string theory on the near horizon AdS geometry. 
The AdS$_{2}$/CFT$_{1}$ could conceivably 
be the most significant case in that 
all the extremal black holes contain an AdS$_{2}$ factor 
in their near horizon geometry 
\cite{Kunduri:2007vf,Figueras:2008qh}. 
In spite of a lot of interesting works 
\cite{Strominger:1998yg,Maldacena:1998uz,Nakatsu:1998st,Townsend:1998qp,
Spradlin:1999bn,Cadoni:1999ja,Blum:1999pc,NavarroSalas:1999up,
Caldarelli:2000xk,Cadoni:2000gm,
Bellucci:2002va,
Strominger:2003tm,Leiva:2003kd,
Giveon:2004zz,
Azeyanagi:2007bj,
Alishahiha:2008tv,Gupta:2008ki,Hartman:2008dq,Galajinsky:2008ce,Sen:2008vm,Sen:2008yk,
Sen:2011cn,Chamon:2011xk,
MolinaVilaplana:2012xe,Jackiw:2012ur} 
it would be fair to say that 
we have not yet gained a fully satisfactory understanding 
of the correspondence due to the peculiarities of the
AdS$_{2}$/CFT$_{1}$,  
including the fact that only the AdS$_{2}$ has two disconnected boundaries 
and it is a long-standing question 
whether the dual CFT$_{1}$ description 
is a single CFT or two systems living on the two boundaries.

In this letter 
we study conformal quantum mechanics (CQM) without 
specifying classical Lagrangian description. 
One important consequence is 
a generic evidence of the AdS$_{2}$/CFT$_{1}$  
as the relationship between 
a generating function of the dilatation expectation values 
in the boundary CQM and a partition function of the bulk AdS$_{2}$. 
We show that 
the expectation values are not associated with 
the ground state but with two excited states in the correspondence.  
Our result is in favor of the statement 
\cite{Zamolodchikov:2001ah,Sen:2008vm,Balasubramanian:1998sn,Balasubramanian:1998de,Balasubramanian:2010ys} 
that the dual CFT$_{1}$'s on two boundaries 
of AdS$_{2}$ space-time are excited and entangled. 
We also claim that 
the energies on the boundary CQM 
would be responsible for the AdS$_{2}$ radius 
so that the ground state would correspond to a flat space 
in the bulk with an infinitely large AdS$_{2}$ radius. 

Another intriguing thing is  
the speculative implication of the resulting expectation values 
of the dilatation operator in CQM.  
When we consider the DFF-model \cite{deAlfaro:1976je} as CQM, 
the dilatation operator takes the form of $xp$. 
Such operator has been proposed as a strong 
candidate for the realization of the Hilbert-P\'{o}lya conjecture 
that the imaginary part of the non-trivial 
Riemann zeros are eigenvalues of a self-adjoint operator 
(see 
\cite{MR1694895,berry1999h,MR1684543,MR2443603,Sierra:2011tb,MR2812337} 
and references therein). 
Unlike a lot of efforts undertaken thus far 
we here propose a novel approach to obtain the counts of the Riemann zeros 
from CQM point of view. 
The fact that the operator $xp$ is the dilatation operator in CQM rather
than the Hamiltonian enables us to jump into a fairly general setting 
beyond the operator with the form of $xp$. 
We show that 
the expectation values of the dilatation operator in CQM 
naturally lead to the asymptotic form of the smoothed counting function 
of the Riemann zeros.

We shall begin by considering CQM  
that is invariant under the conformal symmetric transformation 
of the time coordinate $t$ \cite{deAlfaro:1976je}
\begin{align}
\label{conf1c}
\delta t&=\epsilon_{1}+\epsilon_{2}t+\epsilon_{3}t^{2}
\end{align}
where $\epsilon_{1}$, $\epsilon_{2}$ 
and $\epsilon_{3}$ are identified with the 
infinitesimal parameters of the translation, the dilatation 
and the conformal boost transformation respectively. 
The corresponding generators, 
the Hamiltonian $H=id/dt$, the dilatation operator
$D=it\left(d/dt\right)$,  
and the conformal boost operator $K=it^{2}\left(d/dt\right)$ 
obey the commutation relations 
\begin{align}
\label{conf1e1}
[H,D]&=iH\\
\label{conf1e2}
[K,D]&=-iK\\
\label{conf1e3}
[H,K]&=2iD
\end{align}
of $SL(2,\mathbb{R})$, 
which we call the one-dimensional conformal group. 
Hence the Hilbert space of conformal quantum mechanical system 
exhibits the $SL(2,\mathbb{R})$ conformal symmetry 
and the physical states would be classified 
by its irreducible representation. 
Since we wish to make all integrals convergent,  
we require the unitarity of the representation. 
The classification of the irreducible unitary representations of 
$SL(2,\mathbb{R})$ was studied in \cite{Bargmann:1946me} 
and the irreducible unitary representation 
of $SL(2,\mathbb{R})$ conformal group 
as a function of the continuous time coordinate $t$ can be
generalized by taking the principal spherical series of the representation 
which is induced by the one-dimensional representation 
of the Borel subgroup. 
Let $V_{\lambda}$ be a set of the irreducible unitary representations 
of $SL(2,\mathbb{R})$ with weight $\lambda$. 
Then the Hamiltonian $H$, the dilatation operator $D$ 
and the conformal boost operator $K$ can be expressed as
\begin{align}
\label{conf1d1}
H&=i\frac{d}{dt}\\
\label{conf1d2}
D&=it\frac{d}{dt}
+\frac{\lambda}{2i}
\\
\label{conf1d3}
K&=it^{2}\frac{d}{dt}
+\frac{1}{i}\lambda t,
\end{align}
satisfying the commutation relations
(\ref{conf1e1}), (\ref{conf1e2}) and (\ref{conf1e3}). 
The unitarity implies that 
$\frac12(\lambda+1)$ is pure imaginary. 
The finite conformal transformation is 
\begin{align}
\label{conf1a}
t^{n}&\rightarrow \frac{(\alpha t+\beta)^{n}}
{(\gamma t+\delta)^{\lambda+n}}
\end{align}
where the parameters $\alpha$, $\beta$, $\gamma$ and $\delta$ 
are the elements of real two by two matrices with determinant one
\begin{align}
\label{conf1b}
A&=\left(
\begin{array}{cc}
\alpha&\gamma\\
\beta&\delta\\
\end{array}
\right),&
\alpha\delta-\beta\gamma&=1
\end{align}
which forms the one-dimensional conformal group $SL(2,\mathbb{R})$. 
The quadratic Casimir operator is given by
\begin{align}
\label{conf1e}
\mathcal{C}_{2}&=HK-iD-D^{2}\nonumber\\
&=\frac{\lambda^{2}}{4}+\frac{\lambda}{2}.
\end{align}
Let us define the ground state $|0\rangle_{\Delta}$ by 
\begin{align}
\label{conf1f1}
H|0\rangle_{\Delta}&=0\\
\label{conf1f2}
D|0\rangle_{\Delta}&=\Delta|0\rangle_{\Delta}.
\end{align}
From the equations (\ref{conf1e})-(\ref{conf1f2}) we see that 
the eigenvalue $\Delta$ of the ground state $|0\rangle_{\Delta}$ 
is $\Delta=\frac{\lambda}{2i}$ 
and we thus write the ground state as $|0\rangle_{\lambda}$. 
Now consider the energy eigenstate obeying 
\begin{align}
\label{conf1f3}
H|E\rangle=E|E\rangle.
\end{align} 
This eigenvector $|E\rangle$ 
is known as the Whittaker vector 
in the representation theory of $SL(2,\mathbb{R})$ 
\cite{MR0271275,MR0311838,MR507800}. 
Given eigenvalue $E$ 
and irreducible representation with weight $\lambda$, 
there is a unique Whittaker vector and we can write 
\begin{align}
\label{conf1g3}
|E\rangle_{\lambda}
&=-\sum_{k}\sum_{n}C_{k}
\frac{(-EK)^{n}}
{n!\lambda(\lambda-1)\cdots(\lambda-n+1)}
|0\rangle_{\lambda,k}
\end{align}
where $k$ parametrizes the degenerate ground states 
which might arise from some symmetries in the theories 
and $C_{k}$ is the weighted coefficient for $k$-th ground state. 
For simplicity let us assume the vanishing of 
tunneling amplitudes; $_{\lambda,l}\langle 0|0\rangle_{\lambda,k}=\delta_{l,k}$ 
and the normalization of the ground states; $\sum_{k}|C_{k}|^{2}=1$ 
so that we shall omit the indices $k$. 
Correspondingly we can take the dual energy eigenstate vector 
$_{\lambda}\langle E|$ as the dual Whittaker vector 
which satisfies the relation 
\footnote{The choice of pair of the Whittaker vector 
and the dual Whittaker vector results in 
the consistent Hamiltonian reduction. }
\begin{align}
\label{conf1h1}
\ _{\lambda}\langle E|K
&=E\ _{\lambda}\langle E|
\end{align}
and it can be represented by
\begin{align}
\label{conf1h2}
\ _{\lambda}\langle E|
:=-_{\lambda}\langle 0|
\sum_{k,n}C_{k}^{*}\frac{(-EH)^{n}}
{n!\lambda(\lambda-1)\cdots (\lambda-n+1)}.
\end{align}

With the Whittaker vector $|E\rangle_{\lambda}$ and 
its dual vector $_{\lambda}\langle E|$ 
as the energy eigenstate vector and its dual vector 
in conformal quantum mechanics, 
we will consider the situation 
where the theory is coupled to two-dimensional bulk theory 
to investigate the AdS$_{2}$/CFT$_{1}$ correspondence. 
Let us consider the function defined by
\begin{align}
\label{conf1h3}
W_{\lambda,E_{L},E_{R}}(\phi)
&:=\ _{\lambda}
\left\langle E_{L}|e^{-2i\phi D}
|E_{R}
\right
\rangle_{\lambda}
\end{align}
where $E_{L}, E_{R}$ are the eigenvalues 
of the Whittaker vector (\ref{conf1f3}) 
and the dual Whittaker vector (\ref{conf1h1}) respectively. 
Here $\phi$ is regarded as the restriction of some bulk field in AdS$_{2}$
space-time on the boundary 
that is coupled to the dilatation operator $D$ 
in conformal quantum mechanics on the boundaries. 
By acting the quadratic Casimir operator (\ref{conf1e}) on 
the function (\ref{conf1h3}), 
one obtains the differential equation
\begin{align}
\label{conf1h4}
\left[
\frac12\frac{\partial^{2}}{\partial\phi^{2}}
-\frac{\partial}{\partial \phi}
-2E_{L}E_{R}e^{2\phi}
\right]W_{\lambda,E_{L},E_{R}}(\phi)
&=
\left(
\frac{\lambda^{2}}{2}+\lambda
\right)W_{\lambda,E_{L},E_{R}}(\phi).
\end{align}
The function $W_{\lambda,E_{L},E_{R}}(\phi)$ 
is known as the $SL(2,\mathbb{R})$ Whittaker function 
\cite{MR0271275,MR0311838,MR507800,MR1729357}. 
At first sight one might expect that 
the Whittaker function $W_{\lambda,E_{L},E_{R}}$ plays a role 
of the generating function of the expectation values of 
the operator $D$. 
However, the unitarity asserts that 
the eigenvalue $\Delta$ of the operator $D$ associated with 
the ground state $|0\rangle_{\lambda}$ are not real-valued observables. 
Alternatively if we consider a shifted operator $\left(D-\frac{i}{2}\right)$, 
the corresponding eigenvalues would provide real-valued physical quantities. 
Furthermore 
in a more precise treatment 
one can express the bulk filed $\phi$ 
as $\beta\phi_{0}$ 
where $\phi_{0}$ is the time-dependent part 
while $\beta$ is the time-independent part. 
Instead of the Whittaker function (\ref{conf1h3}) 
let us consider the function 
\begin{align}
\label{conf1h5}
\Psi_{\lambda,\beta,E_{L},E_{R}}(\phi_{0})
:=\ _{\lambda}\langle E_{L}|
e^{-2i\beta\phi_{0}(D-\frac{i}{2})}
|E_{R}\rangle_{\lambda}.
\end{align}
Applying the quadratic Casimir operator (\ref{conf1e}) on 
the function (\ref{conf1h5}), we get 
\begin{align}
\label{conf1h6}
\left[
\frac12\frac{\partial^{2}}{\partial\phi_{0}^{2}}
-2\beta^{2}E_{L}E_{R}e^{2\beta\phi_{0}}
\right]
\Psi_{\lambda,\beta,E_{L},E_{R}}(\phi_{0})
&=\frac12 
\beta^{2}
\left(\lambda+1\right)^{2}
\Psi_{\lambda,\beta,E_{L},E_{R}}(\phi_{0}).
\end{align}
We see that 
the equation (\ref{conf1h6}) 
is the Wheeler-DeWitt equation 
that is encountered in the minisuperspace approximation 
of Liouville field theory (LFT). 

Local properties of LFT can be described by the Lagrangian 
\begin{align}
\label{lft1a1}
\mathcal{L}&=\frac{1}{4\pi}
\partial^{\mu}\phi \partial_{\mu}\phi
+\mu e^{2b\phi}
\end{align}
where $b$ is the dimensionless coupling constant 
and $\mu$ is the cosmological coupling constant. 
The equation of motion is 
\begin{align}
\label{lft1a2}
\Delta\phi&=4\pi b\mu e^{2b\phi}.
\end{align}
In two dimensions 
it is always possible to make 
any metric $g_{\mu\nu}$ conformally flat by 
coordinate redefinition $g_{\mu\nu}=e^{2b\phi}\eta_{\mu\nu}$. 
Furthermore, in two dimensions, 
the curvature can be determined by the scalar curvature. 
Equation (\ref{lft1a2}) asserts that the curvature 
is a negative constant $-8\pi b^{2}\mu$ and 
$g_{\mu\nu}$ describes a two-dimensional surface with 
constant negative curvature, thus the corresponding Lorentzian surface 
can be identified with AdS$_{2}$ space-time locally. 

In order to quantize LFT via canonical quantization,  
we shall employ the Fourier decomposition of the Liouville field $\phi$ 
and its canonical momentum $\Pi$ on the cylinder
\begin{align}
\label{lft1a3}
\phi(t,\sigma)&
=\phi_{0}(t)+\sum_{n\neq 0}\frac{i}{n}
\left[
a_{n}(t)e^{-in\sigma}+b_{n}(t)e^{in\sigma}
\right]
\nonumber\\
\Pi(t,\sigma)&
=p_{0}(t)+\sum_{n\neq0}
\left[
a_{n}(t)e^{-in\sigma}+b_{n}(t)e^{in\sigma}
\right]
\end{align}
with $a_{n}^{\dag}=a_{-n}$, $b_{n}^{\dag}=b_{-n}$. 
The canonical relation
\begin{align}
\label{lft1a4}
[\phi(t,\sigma),\Pi(t,\sigma')]=i\delta(\sigma-\sigma')
\end{align}
leads to the commutation relations
\begin{align}
\label{lft1a5}
[\phi_{0},p_{0}]&=i, &
[a_{n},a_{m}]&=\frac{n}{2}\delta_{n,-m},&
[b_{n},b_{m}]&=\frac{n}{2}\delta_{n,-m},
\end{align}
which imply that $a_{n}, b_{n}$ are creation operators while 
$a_{-n}, b_{-n}$ are annihilation operators. 
The spectrum of LFT has been discussed 
in the minisuperspace approximation \cite{Seiberg:1990eb}. 
The minisuperspace approximation 
was originally proposed in quantum gravity 
\cite{DeWitt:1967yk,Misner:1972js}
where the problem is simplified by 
only treating the zero mode and 
truncating the higher excited modes. 
Whether physics in minisuperspace quantization 
gives a faithful properties of quantum gravity remains an open question, 
however, it has been discussed that 
the minisuperspace approximation would be exact 
in pure two-dimensional gravity \cite{Moore:1991ir,Fateev:2000ik}. 
Replacing the canonical momentum $p_{0}=\frac{\dot{\phi}_{0}}{2\pi}$ 
with differential operator $-i(\partial/\partial \phi_{0})$, 
we obtain the minisuperspace Schr\"{o}dinger equation
\begin{align}
\label{lft1a6}
\left[
-\frac12 \frac{\partial^{2}}{\partial \phi_{0}^{2}}
+2\pi \mu e^{2b\phi_{0}}
\right]\Psi_{P}(\phi_{0})
&=2P^{2}
\Psi_{P}(\phi_{0})
\end{align}
where $P$ is the Liouville momentum, the eigenvalue of the Hamiltonian 
and $\Psi_{P}(\phi_{0})$ is the minisuperspace wave function. 
For $\phi_{0}\rightarrow -\infty$ 
the interaction is small and the wave function 
is a linear combination of $e^{\pm iP\phi_{0}}$. 
Because of the complete reflection potential at $\phi_{0}\rightarrow 0$, 
$P$ is restricted to be positive so that 
the incoming wave uniquely determines the reflected wave.

The function $\Psi_{\lambda,\beta,E_{L},E_{R}}(\phi_{0})$ 
in CQM plays a role of the generating function of 
the expectation values evaluated among two excited states.  
On the other hand, the Liouville wave function $\Psi_{P}(\phi_{0})$ 
would be regarded as the partition function of AdS$_{2}$, 
which is a function of the boundary values 
in the sense that a partition function generically transforms 
as a wave function under a change of polarization on field space 
specified at a boundary 
\cite{Aganagic:2003qj,KashaniPoor:2006nc,Cheng:2010yw,Beem:2012mb} 
and LFT describes AdS$_{2}$ space-time in the classical solution. 
Therefore we come to the interesting conclusion that 
AdS$_{2}$ bulk mode $\phi$ that behaves as 
the zero-mode $\phi_{0}$ near the boundary 
is dual to a source term $\phi_{0}(D-\frac{i}{2})$ 
in the CQM on the boundary of AdS$_{2}$ via AdS$_{2}$/CFT$_{1}$
correspondence
\begin{align}
\label{ads2cft1a0}
\Psi_{\lambda,\beta,E_{L},E_{R}}(\phi_{0})
&=\left\langle E_{L}
\left|
e^{-2i\beta\phi_{0} \left(D-\frac{i}{2}\right)}
\right|E_{R}
\right\rangle_{\textrm{CQM}}\nonumber\\
&=Z_{\textrm{AdS}_{2}}(\phi|_{\textrm{bdy}}=\phi_{0})
=\Psi_{P}(\phi_{0}).
\end{align}
We observe that the expectation values in the CQM are evaluated 
between two excited states defined by the Whittaker vector (\ref{conf1f3}) 
and the dual Whittaker vector (\ref{conf1h1}) 
in the correspondence (\ref{ads2cft1a0}). 
The two distinct states in the correspondence (\ref{ads2cft1a0})
would would enable us to have two independent dynamical systems. 
This is consistent to the statement  
\cite{Azeyanagi:2007bj,Sen:2008vm,Balasubramanian:2010ys,Chamon:2011xk} that 
unlike higher dimensional cases 
AdS$_{2}$ in the global coordinate has two boundaries and 
the dual conformal field theory of asymptotically AdS$_{2}$ is 
realized as two systems or two copies of CFT$_{1}$ on the two boundaries 
although we cannot exclude the possibility 
of a single CFT$_{1}$ as the dual description. 
The appearance of the excited states could also 
be in favor of the statement that 
in the Lorentzian AdS/CFT operator expectation values in excited CFT states 
differ from vacuum expectation values due to the 
existence of normalizable propagating states in the bulk  
\cite{Balasubramanian:1998sn,Balasubramanian:1998de}. 

The comparison of the two Wheeler-DeWitt equations 
(\ref{conf1h6}) and (\ref{lft1a6}) 
identifies the coupling constant $b$ in LFT 
with the constant parameter $\beta$ in CQM 
and establishes the dictionary of
parameters between the bulk LFT of AdS$_{2}$ and the boundary CQM as
\begin{align}
\label{cor1}
\frac{\pi\mu}{b^{2}}
&=E_{L}E_{R}\\
\label{cor2}
\frac{P^{2}}{b^{2}}
&=-\frac14\left(\lambda+1\right)^{2}
=\left(\Delta-\frac{i}{2}\right)^{2}.
\end{align}
Therefore 
the relations (\ref{cor1}) and (\ref{cor2}) indicate that 
the quantum gravity on AdS$_{2}$ 
can be described by two conformal quantum mechanical systems on 
the boundary with energies $E_{L}$ and $E_{R}$ 
as a holographic principle \cite{'tHooft:1993gx,Susskind:1994vu}. 

The equation (\ref{cor1}) says that 
the excited states with non-vanishing energy eigenvalues 
$E_{L}$, $E_{R}$ in the boundary CQM are needed to realize 
negative constant curvature of AdS$_{2}$ space-time 
which is generated by the non-trivial interaction term 
with finite and non-vanishing parameters $b$, $\mu$ 
in two-dimensional gravity theory. 
It is illustrative to compare our result 
with the analogous statement in the AdS$_{3}$/CFT$_{2}$ 
that the AdS$_{3}$ radius $l_{3}$ is represented by 
the central charge $c$ of the CFT$_{2}$ through 
the Brown-Henneaux relation $c=\frac{3l_{3}}{2G_{3}}$ \cite{Brown:1986nw} 
where $G_{3}$ is the three-dimensional Newton constant. 
Instead of the central charge 
the energies of the states play an important role 
in the AdS$_{2}$/CFT$_{1}$, 
however, the relation between 
the AdS$_{2}$ radius and the energies is 
even more attractive in that 
one of the other parameters in LFT necessarily comes about. 
In terms of the coupling constant $b$ controlling   
the quantum effect in LFT 
we can write the AdS$_{2}$ radius $l_{2}$ as
\begin{align}
\label{cor3}
\frac{1}{\sqrt{E_{L}E_{R}}}&=2b^{2}l_{2}.
\end{align}
The semiclassical analysis in LFT is valid for small $b$, 
and accordingly our formula (\ref{cor3}) would reflect the fact that 
the ground states with vanishing energies in the boundary CQM 
force the AdS$_{2}$ radius to go to infinity 
and the classical AdS$_{2}$ geometry then becomes flat space-time.

We see from the equation (\ref{cor2}) 
that the Liouville momentum $P$ in two-dimensional gravity theory 
corresponds to the conformal dimension of the ground state 
in the dual CQM. 
As we have discussed, 
the unitarity condition in CQM requires that 
$\frac12 (\lambda+1)$ is pure imaginary. 
This is consistent to the fact that 
the Liouville momentum $P$ is real 
in the dual two-dimensional gravity theory.

We would like to emphasize that the correspondence (\ref{ads2cft1a0}) 
and the dictionaries (\ref{cor1}), (\ref{cor2}), (\ref{cor3}) are quite universal 
since we have not specified 
the conformal quantum mechanical systems so far.  
However, if we contain more specific information characterizing 
dynamical properties and symmetries,  
there would be more fruitful statements 
(the GKP-Witten relation \cite{Gubser:1998bc,Witten:1998qj}) 
in the AdS$_{2}$/CFT$_{1}$ as the extension of the relation
(\ref{ads2cft1a0})
\begin{align}
\label{ads2cft1b0}
\left\langle 
e^{h_{0} \mathcal{O}}
\right\rangle_{\textrm{CQM}}
&=Z_{\textrm{AdS}_{2}}(h|_{\textrm{bdy}}=h_{0})
\end{align}
where $h_{0}$ is some function of 
the boundary values for the bulk field $h$ 
while $\mathcal{O}$ is the dual operator in CQM. 
For a non-flat space the left values in the correspondence 
(\ref{ads2cft1b0}) are 
generally presumed to be computed between two excited states 
from the relation (\ref{cor3}). 
It would be interesting that 
there has been proposals for such relation 
associated with the DFF-model in \cite{Chamon:2011xk} 
and with the counting of microstates of BPS extremal black holes 
in \cite{Sen:2008yk,Sen:2008vm}.

We now consider the resulting expectation values
\begin{align}
\label{conf1h7b}
\ _{\lambda}
\langle E_{L}|
\left(
D-\frac{i}{2}
\right)
|E_{R}
\rangle_{\lambda}
&=\frac{i}{2}
\frac{\delta}{\delta \phi}\Psi_{\lambda,E_{L},E_{R}}(\phi)
\Bigl|_{\phi=0}
\end{align}
and its possible application to one of the deepest mathematical problem, 
the Riemann hypothesis. 
The equation (\ref{conf1h6}) has two linearly independent solutions, 
which are known to be cylindric functions \cite{gradshteyn2000table}. 
By requiring the unitarity we can write the solutions as 
\begin{align}
\label{conf1h8}
\Psi_{\lambda,E_{L},E_{R}}(\phi)
&=\frac{1}{i}K_{\lambda+1}
\left(
2\sqrt{E_{L}E_{R}}e^{\phi}
\right)
\end{align}
where $K_{\nu}(z)$ is the Macdonald function. 
We should note that the Macdonald functions 
of purely imaginary order $\lambda+1$ with positive argument are real. 
The prefactor $\frac{1}{i}$ in the generating function (\ref{conf1h8}) 
guarantees the reality condition of the expectation values in CQM. 
Making use of the recurrence relation 
\begin{align}
\label{conf1h9}
K_{\nu-1}(z)+K_{\nu+1}(z)&=-2\frac{d}{dz}K_{\nu}(z)
\end{align}
and the formula (\ref{conf1h7b}) we can write 
the expectation values between the excited states as
\begin{align}
\label{conf1i1}
\ _{\lambda}
\langle E_{L}|
\left(
D-\frac{i}{2}
\right)|E_{R}
\rangle_{\lambda}
&=
-\frac{z}{4}
\left(
K_{\lambda}(z)+K_{\lambda+2}(z)
\right)
\end{align}
where $z=2\sqrt{E_{L}E_{R}}$. 
Let us now consider 
a one particle conformal quantum mechanical model 
known as the DFF-model \cite{deAlfaro:1976je} 
\begin{align}
\label{dff1}
S=\frac12 \int dt \left(
\dot{x}^{2}-\frac{g}{x^{2}}
\right)
\end{align}
with $g$ being a dimensionless coupling constant. 
For the DFF-model the dilatation operator can be expressed as
$D=-\frac12xp+\frac{i}{4}$ where $p$ is the canonical momentum and the equation 
(\ref{conf1i1}) becomes 
\begin{align}
\label{conf1i2}
\ _{\lambda}
\langle E_{L}|
\left(
xp+\frac{i}{2}
\right)|E_{R}
\rangle_{\lambda}
&=\frac{z}{2}
\left(K_{\lambda}(z)
+K_{\lambda+2}(z)
\right).
\end{align}
It is speculated that 
the Riemann zeros would be realized as eigenvalues 
of the operator which takes the form of $xp$ 
\cite{MR1694895,berry1999h,MR1684543} 
as a promising candidate of the Riemann operator in 
the Hilbert-P\'{o}lya conjecture 
whose eigenvalues are the imaginary part of 
the non-trivial Riemann zeros. 
Berry and Keating \cite{berry1999h,MR1684543} identified the operator
$xp$ with the Hamiltonian 
and imposed the conditions $|x|\ge l_{x}$, $|p|\ge l_{p}$ 
so that $l_{x}l_{p}=2\pi \hbar$ in the phase space. 
Then they found that the semiclassical number $N(E)$ of states 
with the energy between $0$ and $E$ is given by  
the area in the phase space divided by the Planck cell $h=2\pi$
\begin{align}
\label{rh1a}
N(E)&=\frac{E}{2\pi}
\left(\log \frac{E}{2\pi}-1\right)+\mathcal{O}(1)
\end{align}
and observed that (\ref{rh1a}) precisely coincides with 
the asymptotics of the smoothed counting function 
of the number of Riemann zeros \cite{MR0466039}. 
Connes \cite{MR1694895} 
introduced the constraints $|x|\le \Lambda$, $|p|\le \Lambda$ 
where $\Lambda$ is a common cutoff 
and counted the number of such semiclassical states as
\begin{align}
\label{rh1b}
N(E)&=\frac{E}{\pi}\log \Lambda
-\frac{E}{2\pi}\left(
\log \frac{E}{2\pi}-1
\right).
\end{align}
He interpreted the counting formula (\ref{rh1b}) 
as missing spectral lines associated to the smooth Riemann zeros 
which arise in the limit $\Lambda\rightarrow \infty$, 
however, it was reinterpreted as 
a finite size correction from a physical system later in \cite{MR2443603}. 
According to these semiclassical proposals of the Hilbert-P\'{o}lya 
conjecture it has been desirable to replace 
these artificially imposed semiclassical regularizations of 
the operator $xp$ with the proper quantum treatment 
which naturally generates a discrete spectrum.  
In order to obtain the discrete spectrum via quantization, 
there has been proposed various attempts including 
the modifications of the $xp$ operator (see, e.g., 
\cite{Sierra:2011tb,MR2812337})
and the adoption of the regularization methods (see, e.g., \cite{MR2443603}). 
Nevertheless, these attempts seem to be quite artificial 
and difficult fo find the quantum mechanical explanation 
to follow these ideas.

Here we wish to propose a novel perspective 
to acquire the distribution of the spectrum of the Riemann operator 
from conformal quantum mechanics point of view. 
Consider now the eigenfunction $\Phi_{\rho}(x)$ of the operator 
$\left(xp+\frac{i}{2}\right)$ in the equation (\ref{conf1i2}) satisfying 
\begin{align}
\label{conf1i3}
\left[
\frac{1}{i}x\frac{d}{dx}+\frac{i}{2}
\right]
\Phi_{\rho}(x)=\rho \Phi_{\rho}(x).
\end{align}
They take the form
\begin{align}
\label{conf1i4}
\Phi_{\rho}(x)=Cx^{\frac12+i\rho}
\end{align}
with $C$ a constant of integration. 
The non-trivial zeros of the Riemann zeta function 
$\zeta(s)$ which is conjectured 
to be $s=\frac12+i\rho$, $\rho\in \mathbb{R}$ in the Riemann hypothesis 
appears in the power $x^{s}$ of the eigenfunction $\Phi_{\rho}(x)$.  
The eigenvalues $\rho$ of the operator $\left(xp+\frac{i}{2}\right)$ 
which would be 
the candidates of the Riemann zeros 
can be continuous in the position eigenfunction.  
This is the same situation 
as has been already discussed in many literatures.  

However, in CQM the operator $\left(xp+\frac{i}{2}\right)$ 
should not be recognized as the Hamiltonian 
but rather as the dilatation operator 
whose expectation values can be measured by the energy eigenstates as
\begin{align}
\label{conf1i50}
D(z;\rho)
&=
\frac{z}{2}
\left(
K_{1-i\rho}(z)
+K_{1+i\rho}(z)
\right).
\end{align}
Note that 
the expression (\ref{conf1i50}) can evidently be lifted to 
arbitrary conformal quantum mechanical systems 
by qualifying $\rho$ as the eigenvalue of 
the operator $-2\left(D-\frac{i}{2}\right)$. 
Although almost all quantum approaches so far have tried to identify 
the operator $xp$ with the Hamiltonian 
and simultaneously diagonalize it 
with the position $x$ or the momentum $p$, 
CQM would provide an alternative avenue to the Riemann hypothesis 
as the diagonalization of the dilatation operator. 
We observe that the ground state $|0\rangle_{\lambda}$ is 
the eigenfunction of both 
of the Hamiltonian and the dilatation operator. 
Since the excited states are not eigenvectors of the dilatation operator, 
the limit $E_{L}$, $E_{R}\rightarrow 0$ of the expectation values 
(\ref{conf1i50}) would naturally give rise to 
the eigenfunction of 
the operator $-2\left(D-\frac{i}{2}\right)$ multiplied by 
its eigenvalue $\rho$. 
In other words, the limit in which $z=2\sqrt{E_{L}E_{R}}$ goes to
zero yields the definite eigenvalue $\rho$ 
and the distribution function $D(\rho)$ of the ground state. 
Hence the expectation values (\ref{conf1i50}) are 
in some sense 
the regularized functions which produce 
the distribution of the eigenvalues $\rho$ as 
\begin{align}
\label{conf1i5}
\rho D(\rho)
&=
\lim_{z\rightarrow 0}
\frac{z}{2}
\left(
K_{1-i\rho}(z)
+K_{1+i\rho}(z)
\right)
\end{align}
where we have used the relation 
$\lambda=-(1+i\rho)$ and the formula $K_{\nu}(z)=K_{-\nu}(z)$. 
The asymptotic behavior of the Macdonald function 
\begin{align}
\label{conf1i6}
K_{1+i\rho}(z)
\sim \sqrt{\frac{\pi}{z}}
e^{-\frac{\pi}{2}\rho}
\left(
\frac{2\rho}{ze}
\right)^{i\rho}
\end{align}
for large $\rho$ allows us to write (\ref{conf1i5}) as
\begin{align}
\label{conf1i7}
\rho D(\rho)
&=\lim_{z\rightarrow 0}
\sqrt{\pi z}e^{-\frac{\pi}{2}\rho}
\cos\left[
\rho\ln \left(
\frac{2\rho}{z e}
\right)
\right].
\end{align}
The semiclassical distribution of (\ref{conf1i7}) 
for large $\rho$ is realized when 
the cosine function is at its maximum
\begin{align}
\label{conf1i8}
\cos\left[
\rho\ln \left(
\frac{2\rho}{ze}
\right)
\right]=1
\end{align}
so that 
\begin{align}
\label{conf1i9}
\frac{\rho}{\pi}
\left[
\ln \left(
\frac{\rho}{E_{L}E_{R}}\right)
-1\right]
&=2n,\ \ \  \forall n\in \mathbb{Z}.
\end{align}
Since the expression (\ref{conf1i9}) diverges when 
$E_{L},E_{R}\rightarrow 0$, 
a low energy cutoff is required to make sense of 
the expression (\ref{conf1i9}). 
Let us introduce the cutoff $\Lambda$ such that 
$E_{L}E_{R}=2\pi/\Lambda$. 
Then we obtain the behavior of the large eigenvalues $\rho$ as 
\begin{align}
\label{conf1i10}
N(\rho)&=\frac{\rho}{2\pi}\ln\Lambda 
+\frac{\rho}{2\pi}\left(
\ln \frac{\rho}{2\pi}-1
\right).
\end{align}
Remarkably the first term is 
a continuum in the limit $\Lambda\rightarrow \infty$ 
while the second term leads to 
the asymptotics of the counting function of the Riemann zeros 
as in (\ref{rh1a}) and (\ref{rh1b}). 
It would be interesting to note that 
the equation (\ref{conf1i10}) also 
counts the large conformal dimensions for the ground state in CQM. 
Combining the semiclassical realization (\ref{conf1i10}) 
of the counting Riemann zeros 
with our proposed holographic correspondence (\ref{cor2}) 
would indicate underlying profound relation among 
essential ingredients in number theory, in quantum mechanics and in gravity.

\subsection*{Acknowledgments}
The author would like to thank 
Pei-Ming Ho, Kazuo Hosomichi, Takeo Inami and Dharmesh Jain 
for communications and discussions 
and Yu Nakayama for enlightening comments and remarks 
about Liouville field theory 
and Michael Berry, Jon Keating, 
Paul Townsend and especially Germ\'{a}n Sierra 
for helpful comments and explanations of their works 
on Riemann zeros. 
This work was supported by National Taiwan University 
and the National Center for Theoretical Science (NCTS).

\bibliographystyle{utphys}
\bibliography{ref}

\end{document}